\documentclass[letterpaper,aps,prl,twocolumn,showcaps,superscriptaddress,amsmath,amssymb]{revtex4}
\usepackage{graphicx,color,natbib}

\usepackage{graphicx,color}
\usepackage{amssymb,amsmath}

\begin{document}

\title{Kinetic Crystallisation Instability in Liquids with Short-Ranged Attractions}


\author{C. Patrick Royall}
\affiliation{H.H. Wills Physics Laboratory, Tyndall Avenue, Bristol, BS8 1TL, UK.}
\affiliation{School of Chemistry, University of Bristol, Bristol, BS8 1TS, UK.}
\affiliation{Centre for Nanoscience and Quantum Information, Tyndall Avenue, Bristol, BS8 1FD, UK.}

\begin{abstract}
Liquids in systems with spherically symmetric interactions are not thermodynamically stable when the range of the attraction is reduced sufficiently. However, these metastable liquids have lifetimes long enough that they are readily observable prior to crystallisation. Here we investigate the fate of liquids when the interaction range is reduced dramatically.
Under these conditions, we propose that the liquid becomes \emph{kinetically unstable}, \emph{i.e.} its properties are non-stationary on the timescale of structural relaxation. Using molecular dynamics simulations, we find that in the square well model with range 6\% of the diameter, the liquid crystallises within the timescale of structural relaxation for state points except those so close to criticality that the lengthscale of density fluctuations couples to the length of the simulation box size for typical system sizes. Even very close to criticality, the liquid exhibits significant structural change on the timescale of relaxation.


\end{abstract}

\maketitle


\section{Introduction}
\label{sectionIntroduction}

Many atomic systems typified by the Lennard-Jones model exhibit a temperature range over which the liquid is thermodynamically stable. When the range of the attraction is short 
relative to the molecular size, such as in C$_{60}$, the system has at most only a tiny temperature range where the liquid is stable ~\cite{hagen1993}. However these and related materials exhibit \emph{metastable} liquid states whose lifetime is long on simulation timescales ~\cite{tenwolde1997,haxton2015}. Relative to the particle diameter, even shorter ranged attractions can be obtained with colloid-polymer mixtures where the strength and range of the effective attraction between the colloids can be tuned with the polymer \cite{poon2002,manoharan2015,zhang2013}. Other systems with similar behaviour include weakly stablized colloids~\cite{solomon2001} and those where the suspending liquid induces critical Casimir attractions between two colloids \cite{hertlein2008,bonn2009,buzzacaro2010}. Moreover short-ranged attractive systems form a basic model for proteins ~\cite{zhang2012sm,fusco2016}, whose interaction range can also be ``tuned'', by the addition of ligands, such that the protein liquid can become thermodynamically stable (in the absence of solvent) \cite{perriman2009}.

\begin{figure}
\centering
\includegraphics[width=80 mm]{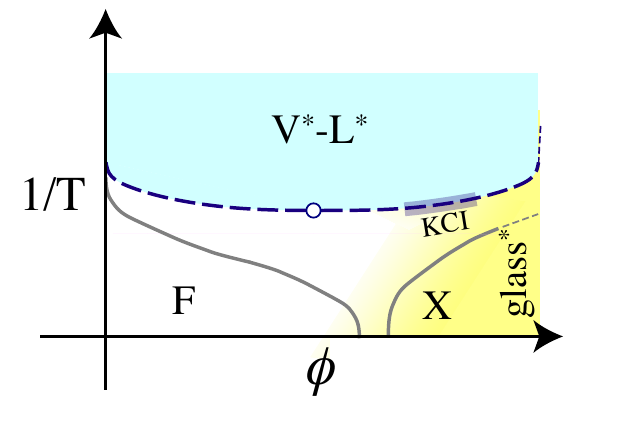}
\caption{Color online. Schematic phase diagram of spheres with a short-ranged attraction in the volume fraction-inverse temperature plane. Shown are thermodynamically stable states, fluid ($F$) and crystalline solid ($X$). Also shown are states which are not thermodynamically stable. These are indicated with an asterisk $^*$ and comprise a vapour ($V^*$), liquid ($L^*$) and glass. Solid grey lines denote thermodynamically stable fluid-crystal ($F-X$) coexistence. Long dashed blue line denotes vapour-liquid coexistence ($V^*-L^*$) which is not thermodynamically stable. The critical point is indicated by the unfilled circle. For spheres with very short-ranged attractions, the spinodal line is very close to the binodal, and is not shown here. Except very close to criticality, the liquid is dense enough to be glassy, which leads to arrested phase separation \emph{i.e.} gelation.  At high densities, spheres undergo dynamical arrest (vitrification) in a continuous fashion, which is indicated by the yellow shaded regions ~\cite{berthier2011,royall2018}. Such slow dynamics hamper the accurate determination of phase boundaries and so in this glassy regime, we give an \emph{indication} of possible phase boundaries by thin dashed lines. In this work, we are interested in a \emph{kinetic crystallisation instability}. To explore this phenomenon, we work along the  ($V^*-L^*$) coexistence line, as indicated by the grey shading marked KCI.
\label{figSchematic}}
\end{figure}

Systems with short-ranged attraction such as colloids can often undergo gelation ~\cite{poon2002,zaccarelli2007,lu2008,zaccarelli2008,nguyen2013} to form a bicontinuous network which can be locally crystalline ~\cite{royall2012,sabin2012,zhang2012,manoharan2015,tsurusawa2017}. Gelation is associated with spinodal liquid-vapour demixing to a bicontinuous network ~\cite{verhaegh1997,lu2008,zaccarelli2008,royall2018,richard2018} where the liquid is dense enough to undergo dynamical arrest leading to the solid-like nature of the gel ~\cite{lu2008,zaccarelli2008,testard2011,zhang2013,testard2014,royall2018}. Gelation in this context corresponds to a state which is intrinsically out-of-equilibrium, due to its dynamical arrest and is illustrated in the schematic phase diagram in Fig. \ref{figSchematic} where gels are found within the vapour-liquid coexistence region.

When gelation is avoided, \emph{i.e.} that the liquid is of sufficiently low density to remain mobile, or the system remains outside the liquid-vapour coexistence region of the phase diagram, density fluctuations in the vicinity of a critical point can massively enhance nucleation rates ~\cite{tenwolde1997,savage2009,fortini2008,babu2009}. Now the density of the liquid in coexistence with its vapour is influenced by (at least) two factors. One is the temperature: approaching criticality, the density of the liquid approaches that of the critical isochore, while upon deeper quenching, the density of such spherically symmetric systems increases. The second factor is \emph{interaction range}. As noted above, longer-ranged systems such as the Lennard-Jones model exhibit a wide range of temperature where the liquid is stable, shorter ranged such as $C_{60}$ feature at best a much reduced range and shorter-ranged interactions again (such as the square well with range 3\%, Eq. \ref{eqSW}) have no thermodynamically stable liquid and undergo gelation upon very weak quenches below the critical temperature ~\cite{royall2018}. Since spheres crystallise at higher density, we infer that a larger temperature range of thermodynamically stable liquid implies a lower density.

We now consider previous work relating to this suggestion of a higher liquid density for short-range interactions in a little more detail. For relatively long interaction ranges, the liquid density increases (for a given degree of cooling with respect to criticality) ~\cite{elliot1999,delrio2002,liu2005,kiselev2006}. For shorter interaction ranges, problems with crystallisation of the liquids (precisely the issue we address here) mean that it can be necessary to use theoretical treatments, such as integral equation theory such as the Self-Consistent Ornstein-Zernike Approximation (SCOZA) to determine the density of liquids in short-ranged attractive systems. Such calculations show that upon decreasing the range of the interaction, the density of the liquid increases ~\cite{foffi2002,elmendouba2008,pini2011}. More recently, simulations in which crystallisation has been suppressed, have confirmed this feature for the square well interaction with range 3\%, and obtained liquids with volume fractions of $\phi=0.59$, well above the freezing transition of, for example, hard spheres ~\cite{royall2018}.

It is important to note that gelation in these systems is related to \emph{spinodal} liquid-vapour demixing ~\cite{lu2008,zaccarelli2008,testard2011,zhang2013,testard2014,royall2018}. But the \emph{binodal} of course defines the liquid-vapour coexistence. However, at least for systems such as the square well with 3\% range (see Eq. \ref{eqSW}), and colloid-polymer mixtures with comparable range, for practical purposes, the spinodal and binodal are almost indistinguishable and the liquid-vapour coexistence line in the phase diagram is rather flat so one finds gelation upon quenching below criticality across a wide range of density (see Fig. \ref{figSchematic}). Only very close to criticality is the liquid of sufficiently low density to demix \cite{royall2018}.

In addition to dynamical arrest, characterised by the large (but \emph{continuous} ~\cite{berthier2011,royall2018}) increase in structural relaxation time, $\tau_\alpha$ ~\cite{cavagna2009,berthier2011}, spheres at high density exhibit an important phenomenon for our purposes: relative to the structural relaxation time, the crystallisation time for a given system size becomes very small ~\cite{zaccarelli2009,sanz2010,taffs2013,sanz2014,yanagishima2017}. In particular, in the case of hard spheres, for reasonable system sizes for computer simulation, say $N\sim10^4$ particles, when the volume fraction exceeds around $\phi  \sim 0.57$ the time to crystallise falls below the structural relaxation time ~\cite{taffs2013}. By ``crystallisation time'' we mean that the system is no longer in a stationary state: the fraction of the system identified as crystal by suitable order parameters increases irreversibly ~\cite{taffs2013}. We emphasise that this crystallisation time observation is of course related to system size: for a sufficiently large system, there will be nucleation events arbitrarily close to the phase boundary, and so any thermodynamically metastable hard sphere fluid will nucleate on a short timescale. However, even larger \emph{experimental} systems often struggle to see nucleation significantly closer to the phase boundary than do brute-force computer simulations ~\cite{taffs2013}. Indeed few experiments on hard spheres succeed in observing crystallisation below a volume fraction of $\phi=0.52$ ~\cite{palberg2014}, \emph{i.e.} at a relative increase in density of some 5\% compared to the freezing volume fraction around $\phi=0.492$. These considerations mean that, while the volume fraction quoted above, $\phi  \sim 0.57$, is in no sense an absolute quantity and must of course depend on system size, nevertheless, for system sizes typically encountered, obtaining equilibrated fluid state points for $\phi  \lesssim 0.57$ is usually straightforward, but for higher volume fractions, crystallisation intervenes in the case of monodisperse systems. We thus conclude that hard spheres exhibit a \emph{kinetic crystallisation instability}, at sufficient volume fraction, that is to say the time to crystallise falls below the structural relaxation time.

\emph{Hypothesis --- Kinetically Unstable Liquids.} The considerations we have outlined above lead us to pose the following question. If we accept that the liquid state becomes denser upon shortening the range of the interaction, we expect that two things will occur: the dynamics of the liquid along the binodal will slow, (Fig. \ref{figSchematic}, ``KCI'') and the crystallisation time will fall. Can it be that the crystallisation time of the liquid actually falls relative to its relaxation time so much that it cannot relax? This would suggest that, rather than being thermodynamically \emph{metastable}, as is the case with C$_{60}$ for example ~\cite{hagen1993}, and in fact we may regard the liquid as being \emph{kinetically unstable}. This would mean that while thermodynamically stable liquids are found for longer interaction ranges, one expects that somewhat shorter ranged systems will exhibit long-lived metastable liquids, but that very short-ranged systems such as those formed by colloid-polymer mixtures will crystallise before the liquid can relax.  We recall that nearly identical behaviour should be expected for a range of systems with short-ranged attractions ~\cite{noro2000,foffi2005}. The aim of this paper is to explore this hypothesis, which turns out to be correct.

To test this hypothesis, our strategy is as follows. First we estimate the density the liquid would have were it stable. Remarkably, we encounter some very dense liquids. We then determine the structural relaxation time in those liquids. Finally we investigate the stability of the liquid \emph{i.e.} the time it takes to crystallise. We find that, even very close to criticality, 
the system is unstable to crystallization. Even closer to criticality we find the structural properties of the liquid are non-stationary on the timescale of structural relaxation. In our discussion we note the relevance of our findings in the context of the recent controversy in water ~\cite{holten2012,sciortino2011,palmer2013,limmer2011,limmer2013,palmer2014,chandler2014}.

Before proceeding we consider what is meant by such a kinetically unstable rather than a (thermodynamically) metastable liquid. Clearly, the kinetically unstable liquids are, like the metastable liquids, not thermodynamically stable. To identify kinetically unstable liquids, we shall use the working definition that significant change in the liquid must take place on the timescale of the structural relaxation time $\tau_\alpha$. We will monitor structural properties so the extreme case is that the liquid crystallises, but we shall also consider other measures \emph{i.e.} structural changes while the system nevertheless remains amorphous. We make the arbitrary criterion that 10\% of the system must be identified as crystalline for ``significant'' crystallization to occur. We observe that once 10\% of the system is identified as crystalline, apart from small fluctuations, the fraction of crystal always increases, so our findings are qualitatively insensitive to reasonable changes to this threshold value of 10\%.

\begin{figure*}
\centering
\includegraphics[width=130 mm]{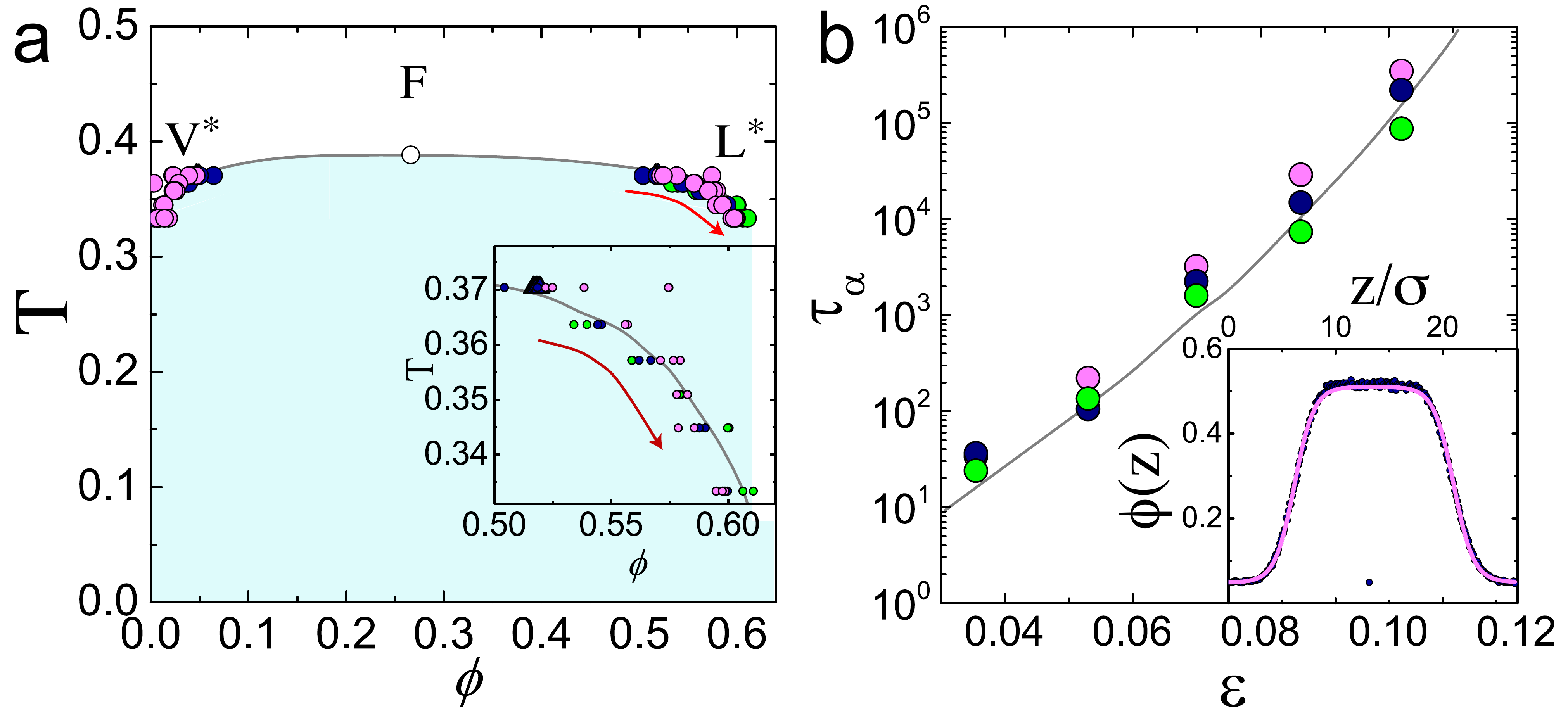}
\caption{
(a) Liquid-vapour binodal for the square well potential with range 6\%. Shown are fluid (F), vapour ($V^*$) and liquid ($L^*$) phases (see Fig. \ref{figSchematic} for a discussion of these phases).
Here the critical point \cite{largo2008} is denoted as a white circle. 
Coexistence is determined from measurements of densities in simulations where phase separation has completed (see text for details). Polydispersity is used to suppress crystallisation, and here is varied from 8\% (light cyan) to 12\% (dark blue) and 16\% (green).
Two system sizes of $N=10976$ (triangles) and $N=4000$ (circles) are shown. Inset shows data on the liquid branch. Solid lines are estimates of phase boundaries.
(b) Angell plot of structural relaxation time for various polydispersities \emph{along the binodal.} Since the data are plotted for state points along the binodal, temperature is the control parameter, but these correspond to different densities (a). Solid line is a fit using the Volgel-Fulcher-Tamman expression, Eq. \ref{eqVFT}.
Inset: Example fit to obtain coexisting liquid and vapour packing fractions. Solid line is the hyperbolic tangent fit described in the text and points are data from $N=10976$ and $\delta=$12\% for a temperature $T=0.3703, \varepsilon=0.0355$. 
\label{figPhaseAngell}}
\end{figure*}

\section{Simulation Methods} 
\label{sectionSimulationMethods}

Throughout we employ the DynamO event driven molecular dynamics package ~\cite{bannerman2011}. We consider the square well model 

\begin{equation} 
\beta u_\mathrm{SW}(r)=
\begin{cases}
\infty  & \mathrm{for}  \hspace{12pt} r<\sigma \\
- \beta \varepsilon _\mathrm{sw}  &  \mathrm{for} \hspace{12pt} \sigma \leq r < \sigma(1+q_\mathrm{sw} )\\
0 & \mathrm{for } \hspace{12pt}  \sigma(1+q_\mathrm{sw} ) \leq r \\
\end{cases} 
\label{eqSW}
\end{equation}

\noindent 
where $\beta=1/k_BT$, $\sigma$ is the diameter $\varepsilon _\mathrm{sw}$ is the well depth and $q_\mathrm{sw}$ is the interaction range. In our system, we consider five equimolar species of identical mass. This enables us to mimic a polydisperse system. The polydispersity $\delta$ we tune to suppress crystallization as follows (in units of $\sigma$):  $\delta=0.08: \{0.888, 0.9573, 1.0, 1.043, 1.112\}$;    $\delta=0.12:\{0.832, 0.936, 1.0, 1.064, 1.168\}$;  $\delta=0.16:\{0.776, 0.9147, 1.0, 1.085,  1.224\}$. Our unit of time is $\sqrt{m\sigma^2/k_BT}$ where $m$ is the mass of each particle . We set $k_BT=1$ ~\cite{bannerman2011}.

System sizes varied from $N=1372$ to $N=10976$ as described. We fix the interaction range of the square well potential at $q_\mathrm{sw}$ of the mean of two interacting particle diameters. We estimate the critical temperature using the results of Largo \emph{et al.} ~\cite{largo2008} and interpolate their values to our choice of $q_\mathrm{sw}=0.06$. This gives a critical temperature $T^c\approx0.384$ and critical volume fraction $\phi^c=0.2662$. We characterise the proximity to criticality with the reduced temperature

\begin{equation}
\varepsilon=\frac{|T-T^c|}{T^c}.
\label{eqEpsilon}
\end{equation}

To calculate the liquid-vapour binodal we set $N=4000$ and checked some state points with a larger system size of $N=10976$. For determination of the relaxation time $\tau_\alpha$ we set $N=1372$  and equilibrated for at least $10\tau_\alpha$ in the NVT ensemble before sampling for at least a further 10$\tau_\alpha$ in the NVE ensemble, except for the deepest quenches ($T=0.333$, $\varepsilon=0.132$) where we equilibrate for $5\times 10^5$ time units and sample for a further $5\times 10^5$ time units. In the case of the crystallization of monodisperse systems, $N=10976$, except for the deepest quench where $N=1372$. Our choice of ensemble is here motived by work by Berthier \emph{et al.} \cite{berthier2007jcpi,berthier2007jcpii} and is common practise for simulations of supercooled liquids \cite{malins2013jcp,jenkinson2017}.

To analyze the local structure and detect crystallization, we identify the bond network using the Voronoi construction with a maximum bond length of $1.4\sigma$. Having identified the bond network, we use the Topological Cluster Classification (TCC) to decompose the system into a set of local structures which include FCC and HCP local crystalline environments of 12 neighbours around a central particle  ~\cite{malins2013tcc}. The amorphous local structures we identify in addition to the crystalline environments are minimum energy clusters for potentials of varying range ~\cite{doye1995}, which include many topologies in the limit of vanishingly small interaction range ~\cite{arkus2009}.

\section{Results} 
\label{sectionResults}

Our strategy to investigate the lifetime of the liquid is to determine the binodal the system would have in the absence of crystallization. To do this, we consider liquid-vapour phase separation in a weakly polydisperse system. The polydispersity is chosen to be sufficient to suppress freezing, and is varied so that we have some idea of any change in the binodal related to the polydispersity itself. This turns out to be small, so we choose to treat the polydisperse system as if it were the monodisperse system of interest if the liquid were to be stable. We then calculate the relaxation time of the liquid along the binodal. Finally we turn to the monodisperse system and consider its freezing kinetics.

\subsection{Estimating the liquid-vapour binodal}
\label{sectionEstimating}

To determine the binodal we use a polydisperse system to prevent crystallization. In our five-component system, small systems $(N\lesssim500)$ can exhibit fluctuations to crystalline states ~\cite{campo2016}, but for the system sizes we consider here we have never observed crystallization when the polydispersity $\delta \geq 0.08$ ~\cite{royall2018}.

Our aim is to obtain the binodal of the coexisting liquid and vapour and to proceed we follow Godonoga \emph{et al.} \cite{godonoga2010}. In particular, we simulate a system close to the critical isochore and wait for it to demix. To minimise its free energy, the system forms a slab of liquid and a slab of vapour. We can then obtain the coexisting volume fractions directly by fitting a hyperbolic tangent \cite{onuki,godonoga2010,royall2018} which approximately profile across the liquid-vapour interface as a function of $z$. This reads
\begin{equation}
\phi(z) = \phi_0 + \Delta \phi \tanh \left( \frac{z-z_0}{\xi} \right)
\label{eqTanh}
\end{equation}
where  $\phi_0+\Delta \phi $ is the volume fraction of the bulk liquid, $\phi_0-\Delta \phi $ is the volume fraction of the vapour,  and $z_0$ is the location of the interface and $\xi$ is the interfacial width. The average volume fraction between the vapour and liquid is then $\phi_0$. Further details may be found in Ref. \cite{godonoga2010}. A typical fit is shown in Fig. \ref{figPhaseAngell}(b) inset. We neglect the effects of partitioning of the composition into each phase. That is, each phase may have different particle size distributions ~\cite{wilding2005}. For the systems we consider, we shall see that the dependence of the binodal upon polydispersity is not too severe.

In this way we construct the liquid-vapour binodal for various polydispersities in Fig. \ref{figPhaseAngell}(a). The first observation is that the binodal is extremely flat. That is to say, even very close to criticality ($\varepsilon=0.0699$) the liquid is of high density ($\phi \approx 0.576$) and therefore would crystallise rapidly were these attractive spheres to behave like hard spheres \cite{zaccarelli2009}. The second observation is that the effect of varying polydispersity is not too significant here. Indeed, only the inset which zooms in on the high-density region shows any effect of polydispersity.

We therefore make the significant assumption that we can neglect the effect of polydispersity upon the liquid-vapour binodal and therefore that our method enables us to estimate the phase boundary that the monodisperse system would have were crystallization not to intervene. This is indicated by the grey line in Fig. \ref{figPhaseAngell}(a). To estimate the liquid packing fraction we take the mean of our measured values. Note that because our use of polydisperse liquids suppresses crystallisation we may access higher liquid volume fractions than previous work~\cite{miller2003}.

\subsection{Dynamics} 
\label{sectionDynamics}

Before we can discuss crystallization times, we need to know the relevant timescale of the system. At these packing fractions, spheres undergo slow dynamics ~\cite{cavagna2009,berthier2011}. To determine the structural relaxation time $\tau_\alpha$ we calculate the intermediate scattering function (ISF) which reads 
\begin{equation}
F(\mathbf{k},t),=\frac{1}{\rho} \langle | \exp \left( {i \mathbf{k.}[\mathbf{r}(t+t') - \mathbf{r}(t')]} \right) | \rangle 
\label{eqISF}
\end{equation}
where $\mathbf{k}=2 \pi \sigma^{-1}$ is a wave-vector taken close to the peak of the static structure factor and the angle brackets indicate averaging over all particles. The location of the particles is given by $\mathbf{r}$. The structural relaxation time is then obtained by fitting a stretched exponential to the ISF
\begin{equation}
F(\mathbf{k},t)=c \exp \left[-(t/\tau_{\alpha})^{b} \right]. 
\label{eqKWW}
\end{equation}
where $b$ is a stretching parameter. Now we cannot fully equilibrate the deeper quenches (for $\varepsilon \gtrsim 0.102$), so the fit should be taken to be approximate only. Upon ageing, $\tau_\alpha$ typically increases ~\cite{berthier2011,elmasri2010,jenkinson2017}. Thus our values may be taken as a lower limit for the relaxation time. We shall see below that for such state points, crystallization proceeds much faster than the relaxation time, so any underestimation of $\tau_\alpha$ has no qualitative effect on our conclusions. We combine our values for $\tau_\alpha$ into the Angell plot shown in Fig. \ref{figPhaseAngell}(b). The dependence on polydispersity is small and the super-Arrhenius behaviour is weak, \emph{i.e.} that our system behaves as a rather strong glassformer.

We note that the data for $\varepsilon \gtrsim 0.06$ are well described by a straight line, indicating an Arrhenius-like behaviour in the relaxation time. This is somewhat surprising because systems with spherically symmetric interactions typically show a super-Arrhenius or fragile behaviour ~\cite{cavagna2009,berthier2011}. However we emphasise that the time-window we access is rather limited. It is quite possible that super-Arrhenius behaviour could be found upon deeper quenching.  We fit 
our data with the Vogel-Fulcher-Tamman (VFT) equation, 
\begin{equation}
\tau_{\alpha}=\tau_{0}\exp\left[ \frac{D}{T-T_{0}} \right].
\label{eqVFT}
\end{equation}
where $T_{0}$ corresponds to an ideal glass transition temperature  ~\cite{cavagna2009,berthier2011}. Other fits are possible ~\cite{hecksler2008}, in particular those from Mode-Coupling Theory (MCT). However while these capture a dynamic range of up to around four decades for ``fragile'' glassformers ~\cite{cavagna2009,berthier2011}, beyond that without some kind of treatment to account for co-operative relaxation, 
MCT fails to capture the dynamics because in practise the system relaxes via mechanisms not accounted for by the theory ~\cite{brambilla2009,berthier2011}. For systems with an Arrhenius-like behaviour, as is the case here, MCT gives a poorer description of the data, only fitting around one decade in time   ~\cite{berthier2007pre}. Here our purpose is merely to use VFT as a fit, rather than to infer any physical interpretation ~\cite{cavagna2009}. We fit the data with fragility parameter $D=3$, $\tau_0=0.01$ and $T_0=0.327$ and neglect any difference in $\tau_\alpha$ as a function of polydispersity. In any case, little variation is seen for the polydispersities considered. 
We thus assume the relaxation time of the monodisperse liquid is given by Eq. \ref{eqVFT}. One dynamical property we have neglected is critical slowing. As Fig. \ref{figPhaseAngell}(b) shows, for our parameters this is not observed, presumably because we do not approach close enough to criticality for such an effect to be significant or that due to the coupling to the simulation box critical fluctuations are suppressed or that it is not too severe at these short wave vectors we consider ($k=2 \pi$ $\sigma^{-1}$).

\subsection{Crystallization Kinetics} 
\label{SectionCrystallization}
\begin{figure*}
\centering
\includegraphics[width=140 mm]{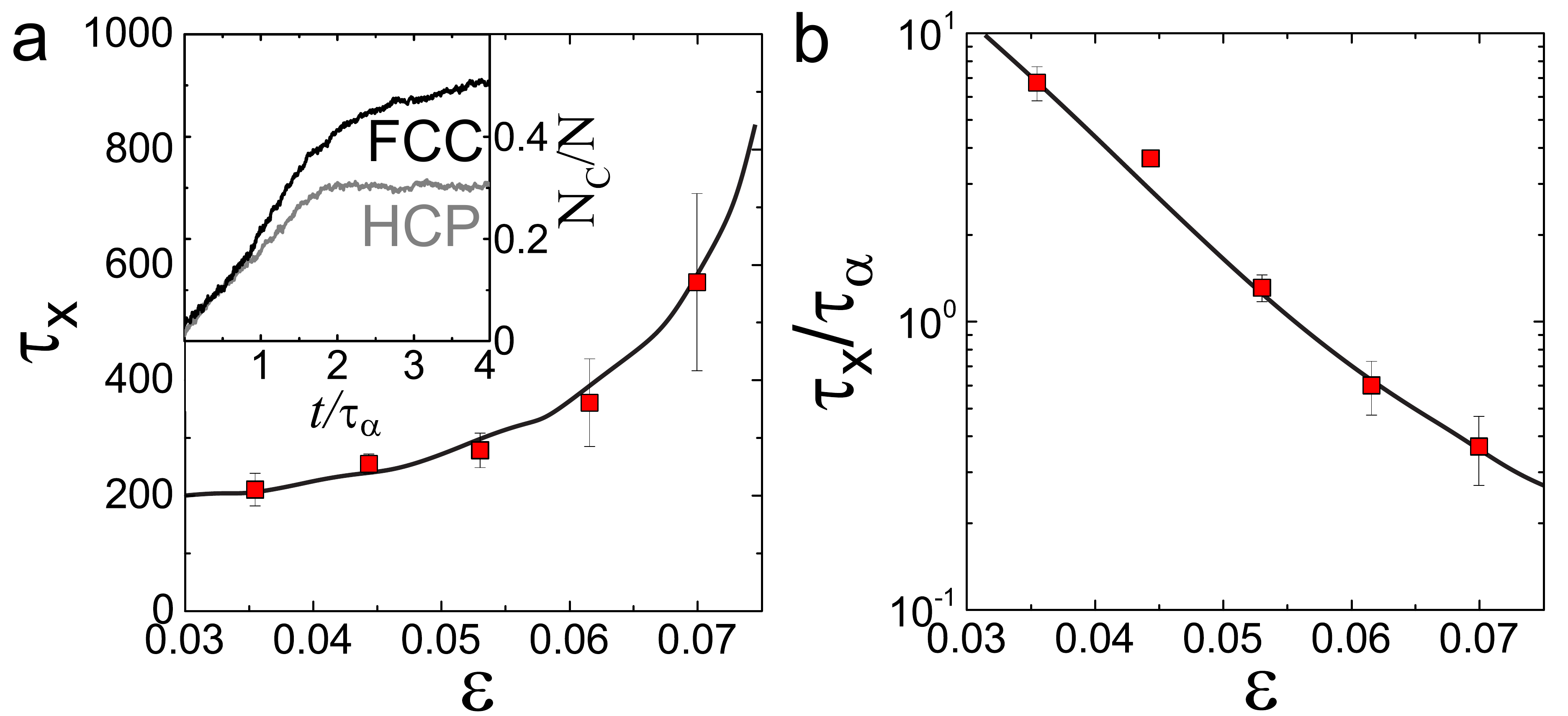}
\caption{Color online. Vanishing liquid stability. (a) Crystallization time as a function of reduced temperature $\varepsilon$ in simulation time units. The solid black line is to guide the eye, dashed grey line in (b) denotes $\tau_\alpha$. Inset: proportion of particles in crystalline environments identified with the TCC as a function of time for a reduced temperature of $\varepsilon=0.0699$. Shown are HCP (grey) and FCC (black) and total crystal fraction. \label{figTxtalTGraf}}
\end{figure*}

Having determined the liquid-vapour phase behaviour and the dynamics, we are now in a position to tackle the hypothesis
with which we opened this article. Is the square well liquid with range $q_\mathrm{sw}=0.06$ kinetically unstable? In Fig. \ref{figTxtalTGraf}(a) inset we show a typical crystallization run for a monodisperse system of $N=10976$ particles at on the binodal reduced temperature $\varepsilon=0.0699$. The system shows  behaviour typical of ``spinodal crystallization'', of continuous growth in the number of particles identified in a crystalline environment, apart from small fluctuations.

We define the crystallization time of the $i$th simulation $\tau_x^{(i)}$ to be when 10\% of the system is identified as being either face centred cubic (FCC) or hexagonal close-packed (HCP). Crystallization is of course stochastic so we average across $N_\mathrm{sim}=6$ independent simulations to define an averaged crystallization time 
\begin{equation}
\tau_x = \frac{  \sum_{i}{N_x}  \tau_x^{(i)} }{N_\mathrm{sim}
}
\label{eqTauXtal}
\end{equation}
where $N_x \leq N_\mathrm{sim}$ is the number of simulations which successfully crystallised. With $\tau_x$ we consider the liquid stability, which is shown in Fig. \ref{figTxtalTGraf}(a). Here we plot as a function of temperature the crystallization time in simulation time units. Now the crystallization time $\tau_x$ increases with cooling. At first sight this is unusual, compared to a typical liquid where it drops as the thermodynamic driving force to crystallise increases with cooling. However the extent of the increase is not startling, less than a factor of five, moreover the density increases markedly as we move along the binodal. Compared to the change in relaxation time [Fig. \ref{figPhaseAngell}(b)], the time to crystallize changes little. We see in Fig. \ref{figTxtalTGraf}(b) that when we scale the crystallization time by the relaxation time, $\tau_x/\tau_\alpha$, the latter dominates strongly. Now we define the liquid to be kinetically unstable where significant change occurs on the timescale of $\tau_\alpha$. Clearly crystallization is a significant change, and reference to Fig. \ref{figTxtalTGraf}(b) shows $\tau_x \leq \tau_\alpha$ is satisfied for $\varepsilon \gtrsim0.045$. Thus only for systems closer to criticality than $\varepsilon \approx 0.045$ might the liquid be considered to be metastable, in the sense that it is expected to last around $\tau_\alpha$ or longer.

\begin{figure}
\centering
\includegraphics[width=85 mm]{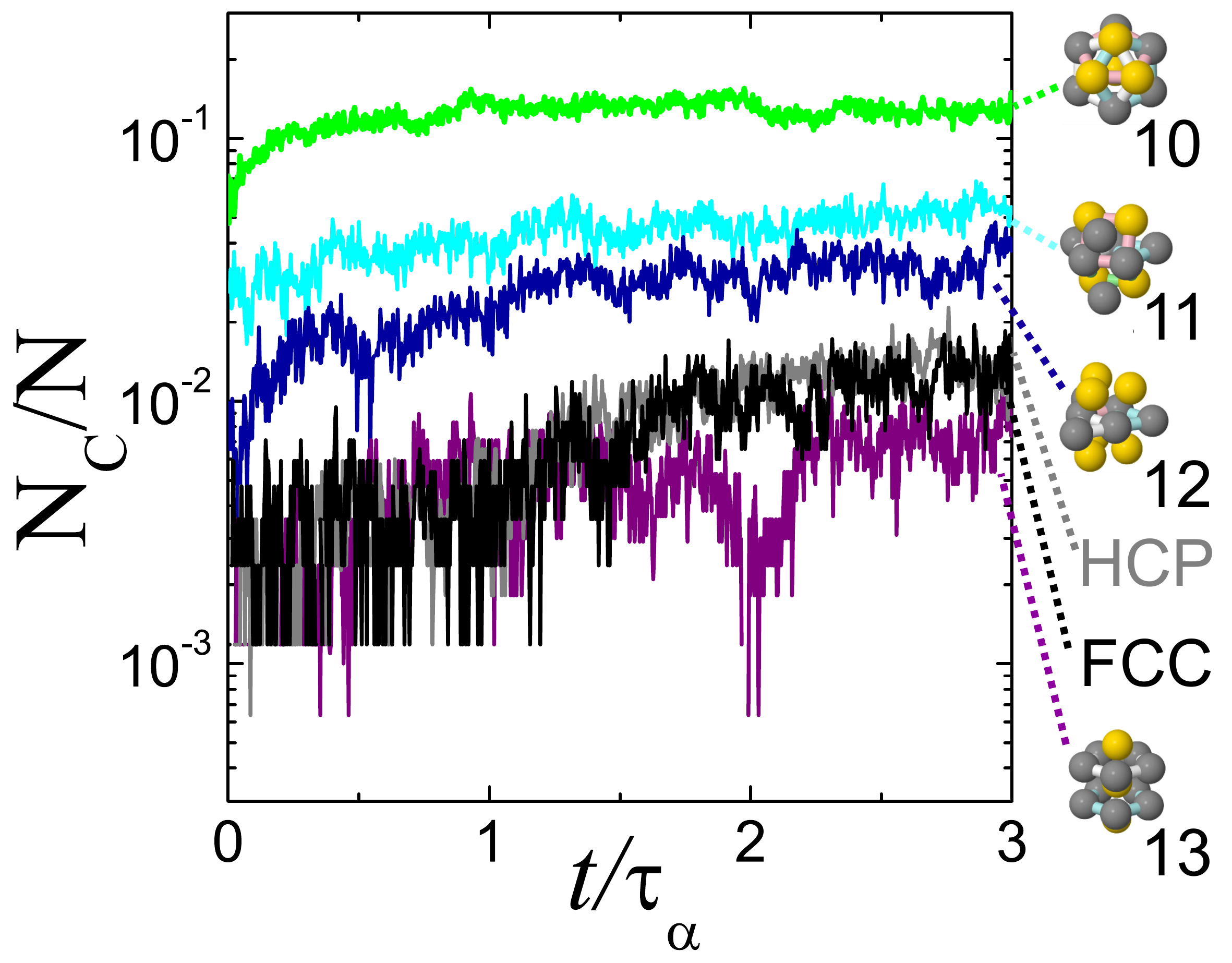}
\caption{Color online. Structural analysis of the liquid at $\varepsilon=0.0355$. Shown are minimum energy clusters with 10, 11, 12 and 13 members, along with FCC and HCP environments.
\label{figStationary}}
\end{figure}

We now consider such a case ($\varepsilon=0.0355$). This is the state point closest to criticality we access and we see in Fig. \ref{figTxtalTGraf}(b) that here the crystallization time $\tau_x=6.74\tau_\alpha$.  Let us probe this state point in more detail to see if it changes on shorter timescales $t \sim \tau_\alpha$. A change in the structure of a non-equilibrium liquid on times greater than the structural relaxation time is not itself surprising: to fully relax a supercooled liquid, it is known that one may need to wait for hundreds of relaxation times \cite{berthier2011,malins2013jcp,jenkinson2017}. To investigate the behaviour of the liquid, we plot amorphous structures identified by the topological cluster classification, in addition to crystalline environments. These local structures are topologically identical to minimum energy clusters for short-ranged Morse potentials with $m=10, 11, 12$ and $13$ particles \cite{doye1995} and are illustrated in Fig. \ref{figStationary}. They consist of a defective icosahedron (a full icosahedron of 13 particles missing three) with $C_{3v}$ symmetry, $m=10$; $C_{2v}$, $m=11$; $D_{3h}$, $m=12$ and a bicapped pentagonal prism ($D_{5h}$ symmetry), $m=13$. We focus on these clusters as they are reasonably prevalent and large enough to provide a sensitive probe of any change in liquid structure than smaller clusters which whose population approaches 100\% in liquids 
\cite{malins2013tcc}.

Considering the population of these clusters, over the timescale of $\tau_\alpha$, we see there is significant change in the population in each case, including the population of particles in local crystalline environments, which increases throughout. Thus we see that, even in the case when the liquid exhibits crystallization on timescales somewhat in excess of its relaxation time, nonetheless it is not in a stationary state prior to crystallization. This is reasonable as to prepare this state point the system is  rapidly compressed from a low density fluid, so some time to equilibrate is expected. We conclude from Fig. \ref{figStationary} that even at this state point very close to criticality, we cannot observe a liquid whose properties are stationary on the relaxation timescale $\tau_\alpha$. Furthermore, while the 10-membered defective icosahedron is incompatible with crystallization due to its fivefold symmetry, crystallization can occur via transient states such as the 12-membered cluster depicted in Fig. \ref{figStationary} \cite{taffs2013}. Its rise in population may thus be a precursor to crystallisation.

\section{Discussion and Conclusions}
\label{sectionDiscussionAndConclusions}

We began this article with the following premis. Upon reducing the range of their attractions, systems with spherically symmetric interactions no longer exhibit a thermodynamically stable liquid state, but the liquids that are found constitute long-lived metastable states. Given that the density of liquids, at a given temperature relative to criticality, increases as the interaction range drops, and noting observations of rapid crystallisation at high density, we argued that there may be an interaction range sufficiently short that the liquid becomes \emph{kinetically unstable} rather than metastable. That is, it is impossible to observe a liquid with stationary properties. We take the structural relaxation time $\tau_\alpha$ as a timescale, and enquire whether the liquid is stationary on that timescale. In particular we find that significant crystallization (10\% of the system in a crystalline environment) takes place on this timescale except very close to criticality with reduced temperature $\varepsilon \lesssim 0.045$.

Interrogating state points even closer to criticality ($\varepsilon=0.0355$), we see that although the liquid lasts marginally longer than $\tau_\alpha$, its structure changes on a timescale of $\tau_\alpha$. We thus conclude that \emph{none} of the state points we sampled exhibits a stationary metastable liquid.  And yet the system considered has a well-defined critical point \cite{largo2008}. Presumably closer to criticality, the time to crystallise would increase, so that it would be possible to access the liquid, and perhaps even the higher-order amorphous structure we have probed would appear stationary. However, simulating such a system even closer to criticality is not trivial. Although finite-size scaling has been developed, enabling a precise mapping onto 3d Ising universality \cite{landau}, this does not reveal all the properties of the system, which may be asymmetric for example. Appealing to brute-force simulation can be problematic, due to divergent lengthscales of the density correlations, so that the structural correlation length can become comparable to the box size. For typical simulation system sizes of $N=10,000$ particles, the box size is of order ten particle diameters, this limits the approach to criticality to $\varepsilon \approx 0.1$.

Experiments on, for example colloid-polymer mixtures are not affected by these concerns. However they are not without their pitfalls. This is due to the necessity to prepare a new sample for each state point which thus far has limited the approach to criticality to $\varepsilon \approx 0.003$ ~\cite{royall2007nphys}. Even innovations to continuously vary the effective temperature via the use of gravity ~\cite{jamie2010} or temperature ~\cite{taylor2012} are hampered by challenges in equilibration close to criticality. Moreover colloids are much more sluggish than molecules and so critical slowing can make equilibration all but impossible on experimental timescales.

Here we have considered the $q_\mathrm{sw}=0.06$ square well, but it is reasonable to suppose that shorter interaction ranges still might be even less kinetically stable. We chose this square well model so that there would be some range where we could determine the structural relaxation time. In the case of denser (less kinetically stable) liquids likely to be encountered with shorter interaction ranges, it would be challenging to quantify $\tau_\alpha$ due to the long relaxation timescales, and thus hard to test for stability.

We close by noting the recent controversy in water, concerning the existence of otherwise of a second liquid \cite{holten2012,sciortino2011,palmer2013,limmer2011,limmer2013,palmer2014,chandler2014}. Our system has a well-defined critical point \cite{largo2008}, but we have shown that even close to criticality ($\varepsilon=0.0355$), the liquid is kinetically unstable. We expect that, very close to the critical point, the liquid might be metastable, and observable on timescales longer than the structural relaxation time. Without entering into a discussion of how easy this might be in the case of water, we suppose that if a critical point can be shown to exist, then it is reasonable to say there is a liquid. However such a quiescent liquid with stationary properties may be impossible to access in simulation or indeed in experiment. Thus the inability to obtain a stationary liquid does not rule out the proximity of a critical point.

\subsection*{Acknowledgements}
It is a pleasure to dedicate this paper to Daan Frenkel, the author would like to thank him for many inspiring discussions over the years.
Zhaleh Ghaemi is gratefully acknowledged for her kind help with some of the preliminary simulations. We would like to thank John Russo and Nigel Wilding for stimulating discussions. Bob Evans is acknowledged for a most critical reading of earlier versions of the manuscript.
CPR gratefully acknowledges the Royal Society, European Research Council (ERC Consolidator Grant NANOPRS, project number 617266) and Kyoto University SPIRITS fund for financial support. This work was carried out using the facilities of the Advanced Computing Research Centre, University of Bristol.


\end{document}